\documentclass[12pt,letterpaper]{article}
\usepackage{hyperref}
\usepackage{times}
\usepackage{amsmath,amsfonts,graphicx}
\usepackage{amssymb}
\usepackage[left=2.54cm,top=2.54cm,right=2.54cm,bottom=2.54cm,bindingoffset=0.0cm]{geometry}
\usepackage{enumitem}
\usepackage{mdframed}
\usepackage{setspace}
\usepackage{titling}
\setlist[enumerate]{topsep=0pt,itemsep=-1ex,partopsep=1ex,parsep=1ex}

\let\OLDthebibliography\thebibliography
\renewcommand\thebibliography[1]{
  \OLDthebibliography{#1}
  \setlength{\parskip}{0pt}
  \setlength{\itemsep}{0pt plus 0.3ex}
}

\usepackage{authblk}

\usepackage{fancyhdr}

\makeatletter
\let\ps@plain\ps@fancy 
\makeatother

\providecommand{\keywords}[1]{\noindent\textbf{\textit{Keywords---}} #1}

\begin{document}
\parskip 0pt

\title{Towards Evaluation of Cultural-scale Claims\\ in Light of Topic Model Sampling Effects}
\author[1,2,*]{Jaimie Murdock}
\author[1]{Jiaan Zeng}
\author[2,3,4]{Colin Allen}
\affil[1]{\footnotesize School of Informatics and Computing, Indiana University, Bloomington, IN 47405, USA}
\affil[2]{\footnotesize Program in Cognitive Science, Indiana University, Bloomington, IN 47405, USA}
\affil[3]{\footnotesize Department of History and Philosophy of Science and Medicine, Indiana University, Bloomington, IN 47405, USA}
\affil[4]{\footnotesize School of Humanities and Social Sciences, Xi'an Jiaotong University, Xi'an, China}
\affil[*]{Corresponding author: \texttt{jammurdo@indiana.edu}}

\date{January 31, 2016}

\maketitle

\bigskip

Cultural-scale models of full text documents are prone to over-interpretation by researchers making unintentionally strong socio-linguistic claims~\cite{Pechenick2015} without recognizing that even large digital libraries are merely samples of all the books ever produced. In this study, we test the sensitivity of the topic models to the sampling process by taking random samples of books in the Hathi Trust Digital Library\footnote{\url{http://hathitrust.org/} --- As of January 31, 2016, the HT consists of nearly 14 million full-text volumes from 89 member libraries.} within different Library of Congress Classification (LCC)\footnote{LCC Outline: \url{https://www.loc.gov/catdir/cpso/lcco/}} areas.

Probabilistic topic modeling has been rapidly adoption in the study of cultural evolution~\cite{Blei2012}. In a topic model, each document is represented as a distribution over topics inferred from the text. Each topic is a probability distribution over words simultaneously inferred from the texts. In the humanities, topic modeling has been used to characterize the evolution of literary diction~\cite{Underwood2012} and of literary studies~\cite{Goldstone2014}. It has also been used to search large corpora for ``the great unread''~\cite{Tangherlini2013}. Topic models have been used to study the large-scale structure of scientific disciplines~\cite{Griffiths2004,Blei2007} and the humanities~\cite{Poetics41, JDH, Jockers2013}. The technique has been deployed as a standard tool in both the JSTOR Data for Research API\footnote{\url{http://about.jstor.org/service/data-for-research}} and the Hathi Trust Research Center (HTRC)'s Data Capsule~\cite{Zeng2014,Murdock2015b}.

The constant addition and revision of works in digital libraries further emphasizes the unrepresentative nature of cultural-scale topic models at any given point in time. If it can be shown that models built from different random samples are highly similar to one another, then researchers can have confidence in their results. In this work, we propose two measures of sampling outcomes.


\textbf{Methods} --- The volumes in four classification areas were downloaded from the HTRC on 19 October 2015 using the HTRC Data API. These areas were selected for their diversity across arts, humanities, sciences, and engineering disciplines. 

For each classification area, we train several topic models over the entire class with different random seeds, generating a set of \emph{spanning models}. Then, we train topic models on random samples of books from the classification area, generating a set of \emph{sample models}. Independent models are trained for each of $k = \{20,40,60,80\}$ topics.

Finally, we perform a topic alignment between each pair of models by computing the Jensen-Shannon distance (JSD) between the word probability distributions for each topic in $\mathcal{M}_1$ and $\mathcal{M}_2$~\cite{Lin1991}. Each topic in $\mathcal{M}_1$ is matched to the closest topic in $\mathcal{M}_2$, allowing for multiple topics in $\mathcal{M}_1$ to be aligned to the same topic in $\mathcal{M}_2$. We take two measures on each model alignment: \emph{alignment distance} and \emph{topic overlap}. The alignment distance is the average JSD of each alignment pair. The topic overlap is the percentage of topics in $\mathcal{M}_2$ that were selected as the nearest neighbor of a topic in $\mathcal{M}_1$.

\begin{figure}[t]
\includegraphics[width=\textwidth]{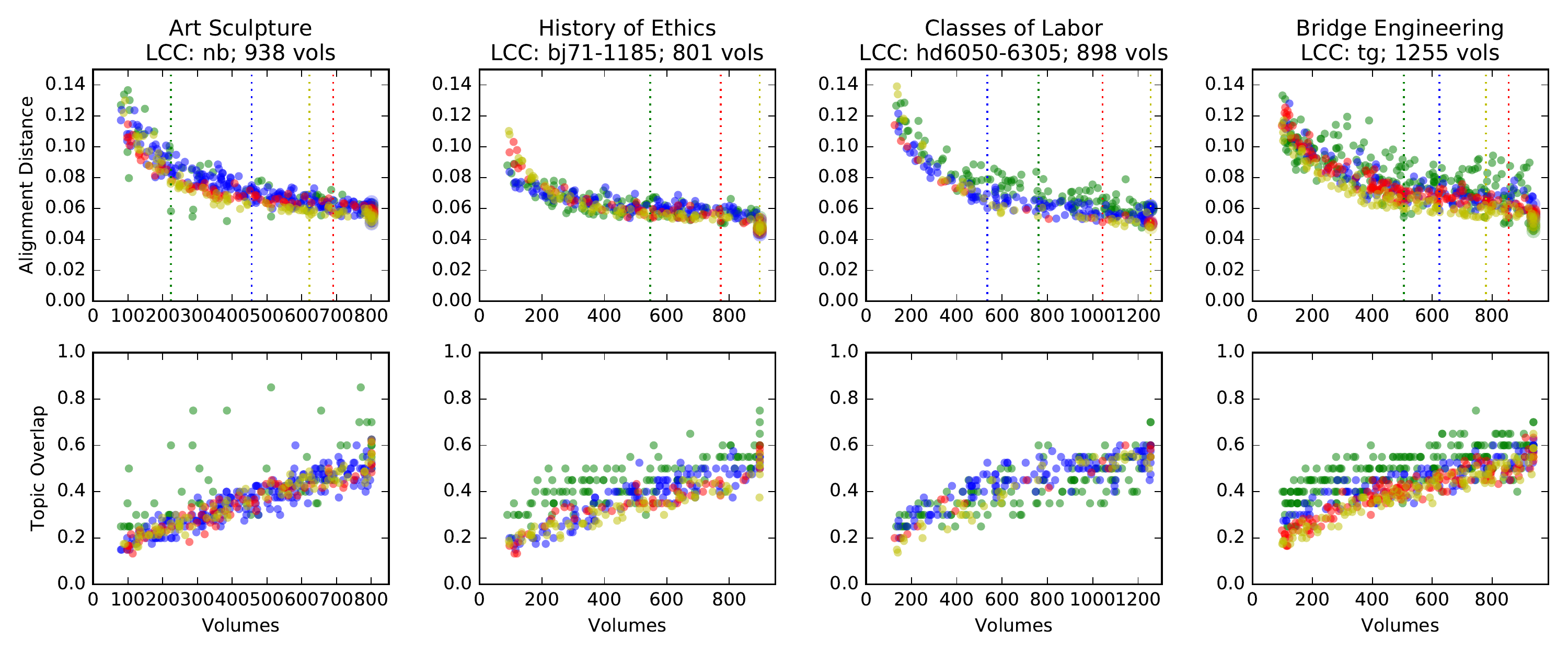}
\caption{\emph{Alignment distance} (top) and \emph{topic overlap} (bottom) for LCC subject area samples in (left-to-right) art sculpture, history of ethics, classes of labor, and bridge engineering. Four topic models are shown: $k=20$ (green), $k=40$ (blue), $k=60$ (red), and $k=80$ (yellow). The dashed lines  in the top charts show the minimum sample size for which alignment distance falls within the variance of alignment distance between spanning models.}
\label{fig:the-figure}
\end{figure}

\textbf{Results} --- We find that sample models with a large sample size typically have an alignment distance that falls in the range of the alignment distance between spanning models (see top row of Fig.~\ref{fig:the-figure}). Unsurprisingly, as sample size increases, alignment distance decreases. We also find that the topic overlap increases as sample size increases. However, the decomposition of these measures by sample size differs by number of topics and by classification area.

\textbf{Conclusion} --- The behavior of different areas may be tied to the ``cognitive extent'' of the discipline~\cite{Milojevic2015}. While this study focuses on only four areas, we speculate that these measures could be used to find classes which have a common ``canon'' discussed among all books in the area, as shown by high topic overlap and low alignment distance even in small sample sizes. For example, discussions of the History of Ethics are likely to discuss Aristotle, Kant, Hume, and Mill regardless of the views championed by the text. 


Our measures of alignment distance and topic overlap provide a content-based evaluation criterion for classification systems, and a validation measure for the robustness of cultural-scale datasets, such as the Google Books corpus~\cite{Michel2011}. Future work is needed to scale these experiments to the entire scope of the LCC in the Hathi Trust.

\section*{Acknowledgments}
The work in this report was supported by a HTRC Advanced Collaborative Support (ACS) Grant.  We thank Miao Chen for her management of the ACS grants. We thank Justin Stamets for assistance with corpus curation.



\bibliographystyle{unsrt}
\bibliography{refs}

\end{document}